\documentclass[aps, prd, onecolumn, tightenlines, notitlepage, superscriptaddress, nofootinbib, preprintnumbers, floatfix,showkeys,11pt,altaffilletter]{revtex4-2}

\usepackage{upgreek}

\usepackage[official]{eurosym}
\usepackage[normalem]{ulem}
\usepackage{amstext}
\usepackage{graphicx}
\graphicspath{{figures/}}
\usepackage{url}
\usepackage{color}
\usepackage{ulem}
\usepackage[version=4]{mhchem}
\usepackage[utf8]{inputenc}
\usepackage{fontawesome}
\usepackage{yfonts}

\pdfoutput=1
\usepackage{textcomp}
\usepackage{comment}
\usepackage{yfonts}
\usepackage{epsfig,amsfonts,mathrsfs,graphicx,color,slashed,multirow}
\usepackage{latexsym,graphicx,slashed,color,enumerate,url,cancel,gensymb}
\usepackage{textcomp}

\usepackage[x11names]{xcolor}
\usepackage[colorlinks,pdfstartview=FitV,breaklinks=true]{hyperref}
\usepackage{booktabs}
\usepackage{adjustbox}
\usepackage{textgreek} 
\usepackage{caption, subcaption} 
\usepackage{booktabs} 
\usepackage{float}

\usepackage{lmodern}
\usepackage{ae,aecompl}
\usepackage{appendix}


\definecolor{vdrgreen}{rgb}{0.0, 0.6, 0.0}

\definecolor{persiangreen}{rgb}{0.0, 0.65, 0.58}
\definecolor{mediumpersianblue}{rgb}{0.0, 0.4, 0.65}

\makeatletter
    \newcommand{\colorboxed}[3][white]{\fcolorbox{#2}{#1}{\m@th$\displaystyle#3$}}
\makeatother

\AtBeginDocument{\hypersetup{citecolor=mediumpersianblue,linkcolor=mediumpersianblue,urlcolor=mediumpersianblue}}
\usepackage{appendix}

\begin{document}

\title{{\Large Addendum to `Combined Analysis of Neutrino Decoherence\\ at Reactor Experiments'}}

\author{Andr\'e de Gouv\^ea}\email{degouvea@northwestern.edu}
\affiliation{Northwestern University, Department of Physics \& Astronomy, 2145 Sheridan Road, Evanston, IL 60208, USA}
\author{Valentina De Romeri}\email{deromeri@ific.uv.es}
\affiliation{Institut de F\'{i}sica Corpuscular CSIC/Universitat de Val\`{e}ncia, Parc Cient\'ific de Paterna\\
 C/ Catedr\'atico Jos\'e Beltr\'an, 2 E-46980 Paterna (Valencia) - Spain}
\author{Christoph A. Ternes}
\email{christoph.ternes@lngs.infn.it}
\affiliation{Istituto Nazionale di Fisica Nucleare (INFN), Laboratori Nazionali del Gran Sasso, 67100 Assergi, L’Aquila (AQ), Italy
}

\begin{abstract}
We update our analyses to constrain neutrino decoherence induced by wave-packet separation with RENO and Daya Bay data, now including the final data sets of the two experiments. We find that while the individual bounds from Daya Bay and RENO data improve relative to our original estimates, the combined fits are still dominated by KamLAND data and are only minimally improved. 
\end{abstract}
\maketitle

\section{Introduction}
In this short note, we update our analyses of RENO and Daya Bay data, presented in Refs.~\cite{deGouvea:2020hfl} and~\cite{deGouvea:2021uvg}, and briefly discuss the consequences on the new inferred bounds on the size of the neutrino wave-packet width obtained from reactor experiments. Here, we make use of the final data releases from both collaborations -- Refs.~\cite{DayaBay:2022orm,RENO_update} -- which correspond to 3158 days of data collection at Daya Bay and 3800 days at RENO. 

In Sec.~\ref{sec:res} we compare the new and old results, and in Sec.~\ref{sec:dis} we discuss the implications of these new results in terms of neutrino wave-packet width\footnote{For the discussion of the effect of a finite wave-packet size on the neutrino oscillation probability we refer the interested reader to the original Refs.~\cite{deGouvea:2020hfl,deGouvea:2021uvg}.}.

\section{Updated analyses and results}
\label{sec:res}

\begin{figure}[t]
\centering
\includegraphics[width=0.48\textwidth]{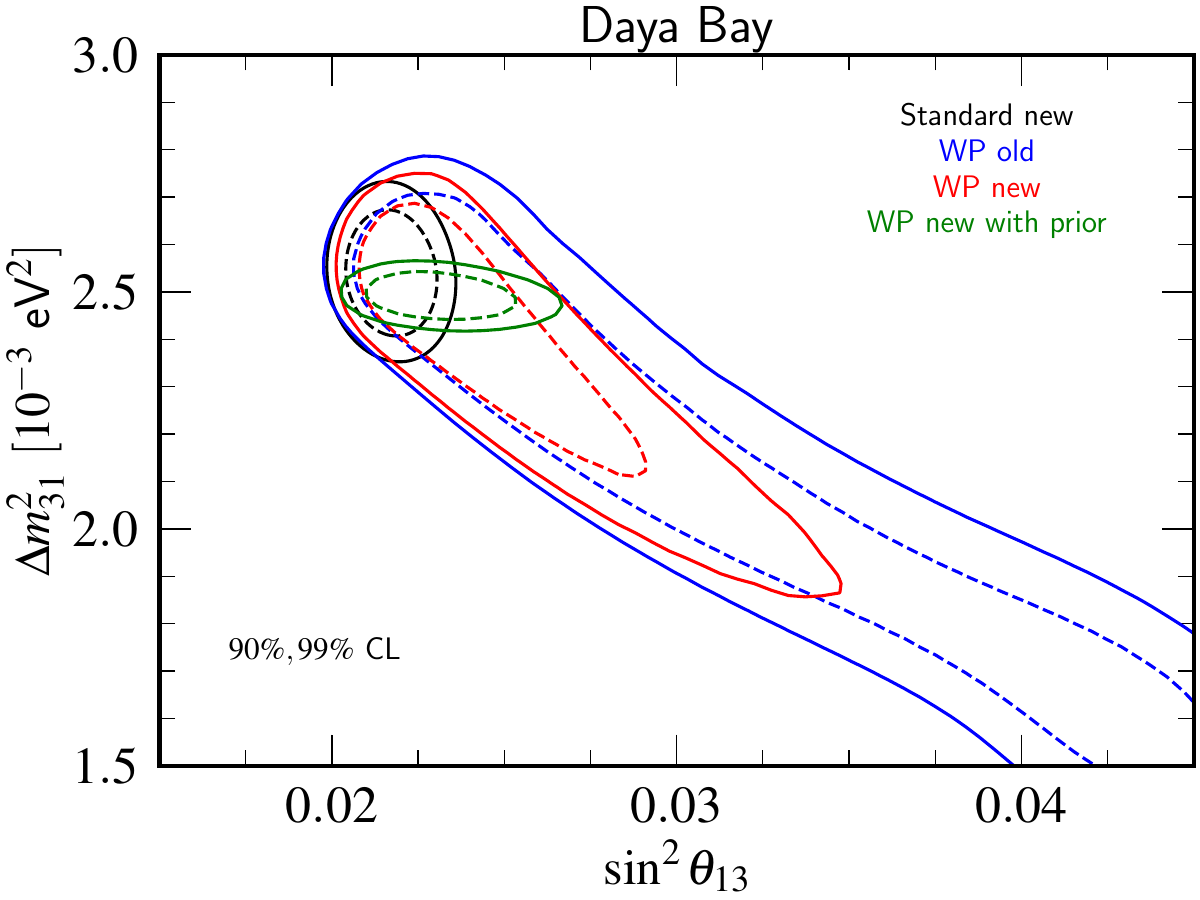}
\includegraphics[width=0.48\textwidth]{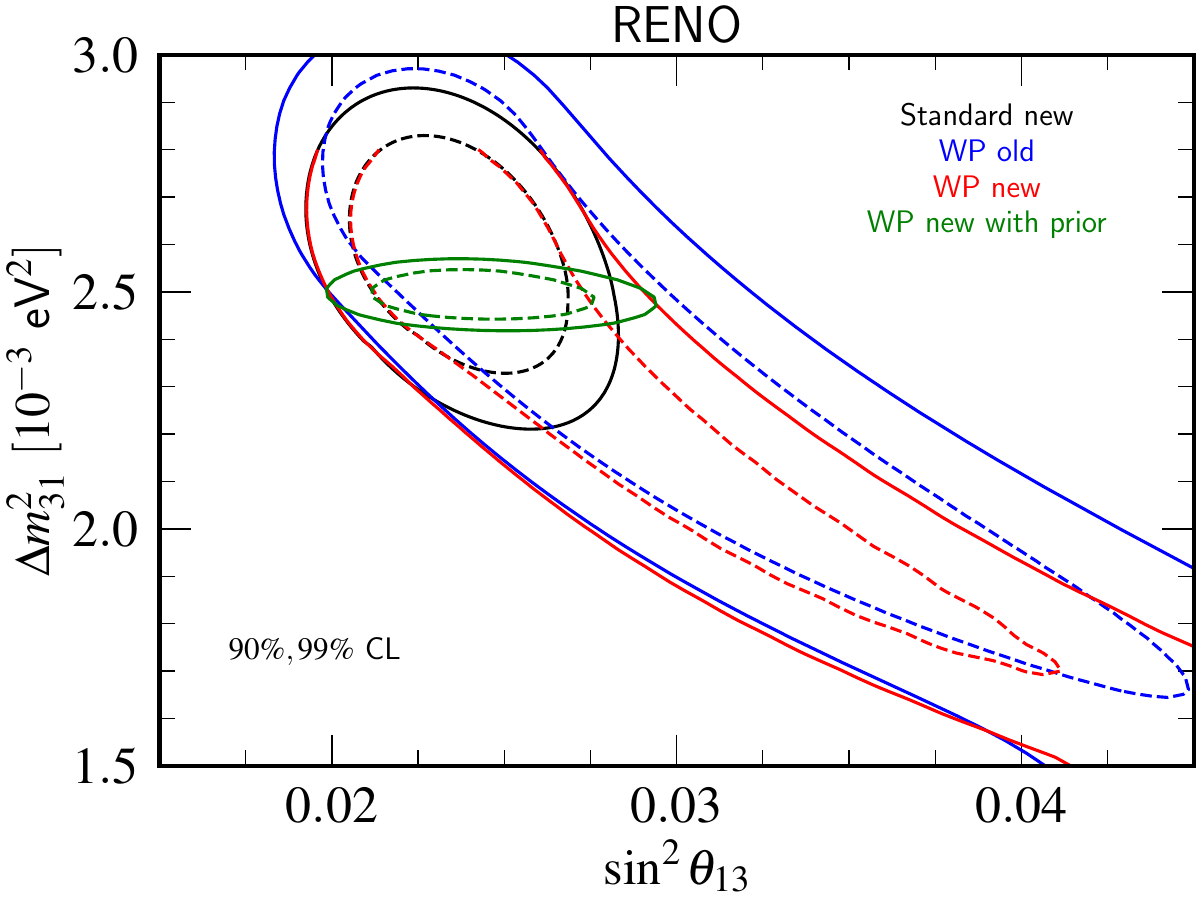}
\caption{90\% CL (dashed) and 99\% CL (solid)  (2 dof) allowed regions in the $\sin^2\theta_{13}-\Delta m_{31}^2$ plane after minimizing over the wave-packet (WP) width for the old data (blue), new data (red) and new data imposing an external prior on $\Delta m_{31}^2$ (green) for Daya Bay (left) and RENO (right). Also shown for comparison is the result from the standard analysis (black) of the newest data.}
\label{fig:comp_both}
\end{figure}

\begin{figure}[t]
\centering    
\includegraphics[width=0.48\textwidth]{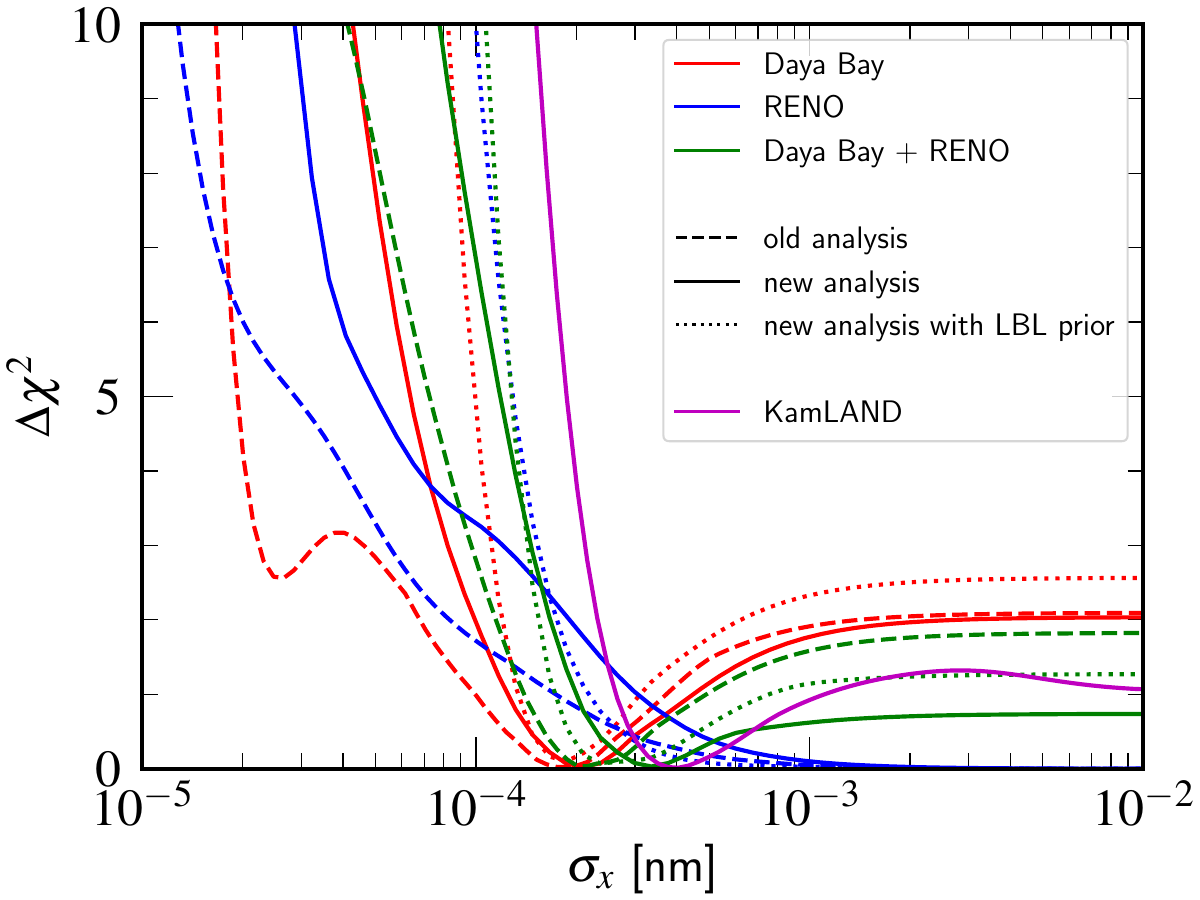}
\includegraphics[width=0.48\textwidth]{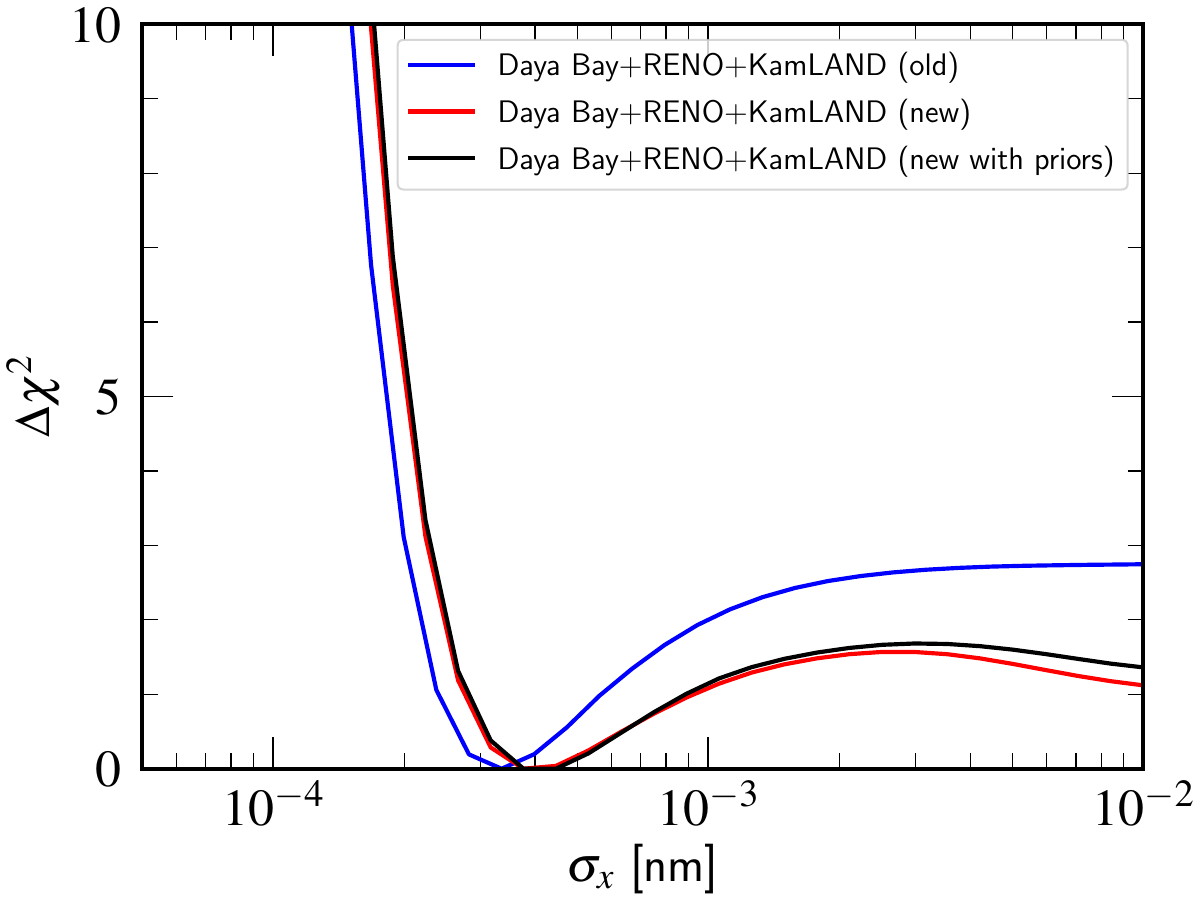}
\caption{$\Delta \chi^2$ profiles obtained from the analyses of reactor data. Left panel: bounds on the neutrino wave-packet width $\sigma_x$ from the analyses of Daya Bay (red), RENO (blue), KamLAND (purple) data and from the combined analysis of RENO and Daya Bay (green). The dashed profiles refer to the old analyses, the solid ones to the analyses of new data and the dotted lines to the new analyses imposing an external prior on $\Delta m_{31}^2$. Right panel: bounds on the neutrino wave-packet width from the combined RENO+Daya Bay+KamLAND analyses with old (blue) and new (red) data sets. The black line refers to the analysis of new data imposing priors on $\Delta m_{31}^2$ and $\sin^2\theta_{12}$.}
\label{fig:comp_sigma}
\end{figure}

We first compare the old and new bounds obtained from the individual analyses of Daya Bay and RENO data. The results are depicted in Fig.~\ref{fig:comp_both}. The left (right) panel contains the allowed regions at 90\% (dashed) and 99\% (solid) confidence level (CL) for 2 degrees of freedom (dof) in the $\sin^2\theta_{13}-\Delta m_{31}^2$ plane, after profiling the size of the neutrino wave-packet $\sigma_x$, obtained from Daya Bay (RENO) data. The blue contours are obtained with the data presented in Refs.~\cite{Adey:2018zwh} and~\cite{jonghee_yoo_2020_4123573}, while the red contours are obtained using the final data releases~\cite{DayaBay:2022orm,RENO_update}. The determination of the oscillation parameters is clearly improved for both experiments. In addition, we perform an analysis imposing a prior on the mass-squared difference -- $\Delta m_{31}^2=(2.490\pm 0.026)\times10^{-3}$~eV$^2$ -- obtained in global analyses of neutrino oscillation data without the inclusion of data from reactor experiments~\cite{deSalas:2020pgw,mariam_tortola_2024_11430808}. This results in a smaller allowed region of the parameter space and affects the extracted bound on $\sigma_x$, as depicted in the left panel of Fig.~\ref{fig:comp_sigma}. From this figure, we see that the bounds obtained from each experiment improve with the new data set, and that the improvement is more pronounced with the prior on $\Delta m_{31}^2$. However, the strongest bounds still come from the analysis of KamLAND data.

In the right panel of Fig.~\ref{fig:comp_sigma} we depict the results of the combined analysis of RENO, Daya Bay and KamLAND data. The blue line is obtained using the previous RENO and Daya Bay data, while the red line is obtained using the final data sets. The bound is only slightly improved with the inclusion of new data. We also show a bound imposing the prior on the mass-squared difference $\Delta m_{31}^2$ and a prior on the solar mixing parameter, $\sin^2\theta_{12}=0.308 \pm 0.020$, obtained from the combined analysis of solar neutrino data~\cite{mariam_tortola_2024_11430808}, which is more precise than the measurement from KamLAND~\cite{Gando:2010aa,kamland_web}. The overall impact on the bound on $\sigma_x$ is again very small. The priors on the oscillation parameters mainly affect the sensitivity at large values of $\sigma_x$ (i.e., in the Standard Model limit). As in previous analyses, we find that the best fit corresponds to a finite wave-packet width with $\sigma_x = 3.75\times10^{-4}$~nm but the overall significance of this non-trivial value is reduced with respect to the previous best-fit value, as is apparent in Fig.~\ref{fig:comp_sigma}.

\section{Discussion}
\label{sec:dis}

We summarize all the updated bounds in Tab.~\ref{tab:bounds}. Some of these exclude the region of parameter space where one could soften the tension in the light sterile neutrino interpretation of the short-baseline anomalies as proposed in Ref.~\cite{Hardin:2022muu}, see also Ref.~\cite{Giunti:2023kyo}. 

Finally, we highlight that a competitive bound on the neutrino wave-packet width was recently presented by the BeEST collaboration~\cite{Smolsky:2024uby}. There, the experimental setup (electron capture) is different from that of reactor experiments, where neutrinos are produced from nuclear beta decay, and therefore the bounds are not directly comparable. Although theoretical expectations of the neutrino wave-packet width are still the subject of debate, they seem to point toward much larger values than those accessible by current experiments of all types~\cite{Akhmedov:2022bjs,Jones:2022hme,Krueger:2023skk}.

\begin{table}[t!]
    \centering
    \begin{tabular}{|c|c|c|}\hline
         \bf{Experiment} & \bf{~~90\% CL~~} & ~~\bf{3$\sigma$}~~ \\\hline
         RENO~\cite{RENO_update} & ~~$1.4\times10^{-4}$~nm~~ & ~~$0.3\times10^{-4}$~nm~~ \\
         Daya Bay~\cite{DayaBay:2022orm} & ~~$0.9\times10^{-4}$~nm~~ & ~~$0.5\times10^{-4}$~nm~~ \\\hline
         RENO~\cite{RENO_update} (with $\Delta m_{31}^2$ prior) & ~~$1.5\times10^{-4}$~nm~~ & ~~$1.0\times10^{-4}$~nm~~ \\
         Daya Bay~\cite{DayaBay:2022orm} (with $\Delta m_{31}^2$ prior) & ~~$1.1\times10^{-4}$~nm~~ & ~~$0.8\times10^{-4}$~nm~~ \\\hline
         RENO~\cite{RENO_update} + Daya Bay~\cite{DayaBay:2022orm} & ~~$1.5\times10^{-4}$~nm~~ & ~~$0.8\times10^{-4}$~nm~~ \\
         RENO~\cite{RENO_update} + Daya Bay~\cite{DayaBay:2022orm} (with $\Delta m_{31}^2$ prior) & ~~$1.4\times10^{-4}$~nm~~ & ~~$1.0\times10^{-4}$~nm~~ \\\hline
         KamLAND~\cite{kamland_web} & ~~$2.2\times10^{-4}$~nm~~ & ~~$1.6\times10^{-4}$~nm~~ \\\hline
         Global & ~~$2.3\times10^{-4}$~nm~~ & ~~$1.7\times10^{-4}$~nm~~ \\
         Global (with $\Delta m_{31}^2$ and $\sin^2\theta_{12}$ priors) & ~~$2.4\times10^{-4}$~nm~~ & ~~$1.8\times10^{-4}$~nm~~ \\
         \hline
    \end{tabular}
    \caption{\label{tab:bounds} Bounds on the size of the neutrino wave-packet width obtained from the analyses of reactor antineutrino data.}
\end{table}

\section*{Acknowledgments}
We thank Sanghoon Jeon and Soo-Bong Kim on behalf of the RENO collaboration for providing us the data presented in Ref.~\cite{RENO_update} in digital form.

The work of AdG is supported in part by the U.S.~Department of Energy grant DE-SC0010143 and in part by the NSF grant PHY-1630782.
VDR acknowledges financial support by the CIDEXG/2022/20 grant funded by Generalitat Valenciana, and from the Spanish grants CNS2023-144124 (MCIN/AEI/10.13039/501100011033 and “Next Generation EU”/PRTR), PID2023-147306NB-I00 and CEX2023-001292-S (MCIU/AEI/10.13039/501100011033). 


\providecommand{\href}[2]{#2}\begingroup\raggedright\endgroup

\end{document}